\documentclass{article}
\usepackage{epsfig}
\usepackage{amsmath}

\title{Exploring Maximum Entropy Distributions with Evolutionary Algorithms}
\author{Raul Rojas\\{Freie Universitaet Berlin} }

\begin{document}

\maketitle

\begin{abstract}
This paper shows how to evolve numerically  the maximum entropy probability distributions for a given set of constraints, which is a variational calculus problem. An evolutionary algorithm can obtain approximations to some well-known analytical results, but is even more flexible and can find distributions for which a closed formula cannot be readily stated. The numerical approach handles distributions over finite intervals. We show that there are two ways of conducting the procedure: by direct optimization of the Lagrangian of the constrained problem, or by optimizing the entropy among the subset of distributions which fulfill the constraints. An incremental evolutionary strategy easily obtains the uniform, the exponential, the Gaussian, the log-normal, the Laplace, among other distributions, once the constrained problem is solved with any of the two methods. Solutions for mixed (``chimera'') distributions can be also found. We explain why many of the distributions are symmetrical and continuous, but some are not.

\end{abstract}

\section{Maximum Entropy Distributions}

The principle of maximum entropy had been used implicitly by statisticians for many years until it became formalized in the mid 1950s. Today, it is used in information theory \cite{cover}, as well as in machine learning \cite{rojas}.
Maximum entropy distributions play an important role in many  applications. Statistical classifiers, for example, try to capture regularities in  data sets keeping a description of the ``data cloud'', which is summarized by a minimal number of parameters. The extreme and opposite case is when the data itself provides its own model, for example in nearest neighbor classifiers (k-NN). In a kNN, new data  is  matched with the closest  point in the data set, and that new point is assigned the class of its nearest neighbor (or $k$ of them, if we decide to classify by taking a majority vote). The other extreme approach is when we summarize a complete cloud of data points by storing only its centroid. Given several classes, each represented by a centroid, we can compute which centroid is  closer to a new data point in order to obtain its classification.

Therefore, given the data, the  problem we generally have is  deciding how many parameters from the ``data cloud'' we want to store (for example, mean value, covariance matrix, and so on). Once we have decided which parameters we want to use, we model the probability  distribution of each class making the least number of assumptions about the shape of the distribution. We apply the principle of ``maximum ignorance'': only the  chosen parameters describe the data set and everything else must be as general as possible. That is, we apply the principle of maximum entropy.

The Gaussian distribution is very popular in this context, because it is well known that given only the mean value and covariance matrix of the data set, the distribution of maximum entropy is a multivariate Gaussian. But there are other probability distributions that can be used: each one of them is the result of applying the principle of maximum entropy to a different set of parameters that we want to store.

In this paper we show, with a few examples, that maximum entropy distributions can be easily found using an evolutionary algorithm that samples from the set of possible probability distributions (defined in a given support interval) constrained by a choice of statistical parameters. The algorithm progresses by selecting distributions with higher and higher entropy, until the search settles on a maximum. This approach is simple but powerful. Discrete distributions can be handled in a straightforward manner. Generally, the maximum entropy property is proved making assumptions about the integrability and differentiability of the distributions, assumptions that do not need to be made in the computational approach illustrated here. We show further down that there are two ways of sampling the space of distributions during the evolutionary optimization.

The continuity and symmetry of some of the maximum entropy distributions that we can find in this way, do not have to be assumed in advance nor have to be enforced during the computation. They arise as emergent properties of the optimal distributions, as we will see further down.

\section{The uniform distribution}

The simplest case we can handle at the beginning is that of a uniform distribution. Given a random variable $X$ which takes real values in the interval $[a,b]$, and if we do not have any further information about $X$, then the most general distribution describing an experiment which produces values of $X$ is the uniform distribution with support in the interval $[a,b]$. It makes sense intuitively.

The same result can be obtained for a discrete distribution $f(x_i)$, for $i=1,…,n$, where each $x_i$ is a discrete value that the random variable $X$ can assume randomly. The entropy of the distribution is defined as
$$
E= - \sum f(x_i) {\rm log}(f(x_i))
$$
Maximizing this function, without constraints, leads to the uniform distribution. The message is that if we do not have any reason to assume that any point $x_i$ is more relevant than any other point, then we should assign each one of them the same probability of being selected.

 \begin{figure}[htb]
\centerline{\includegraphics[width=8cm]{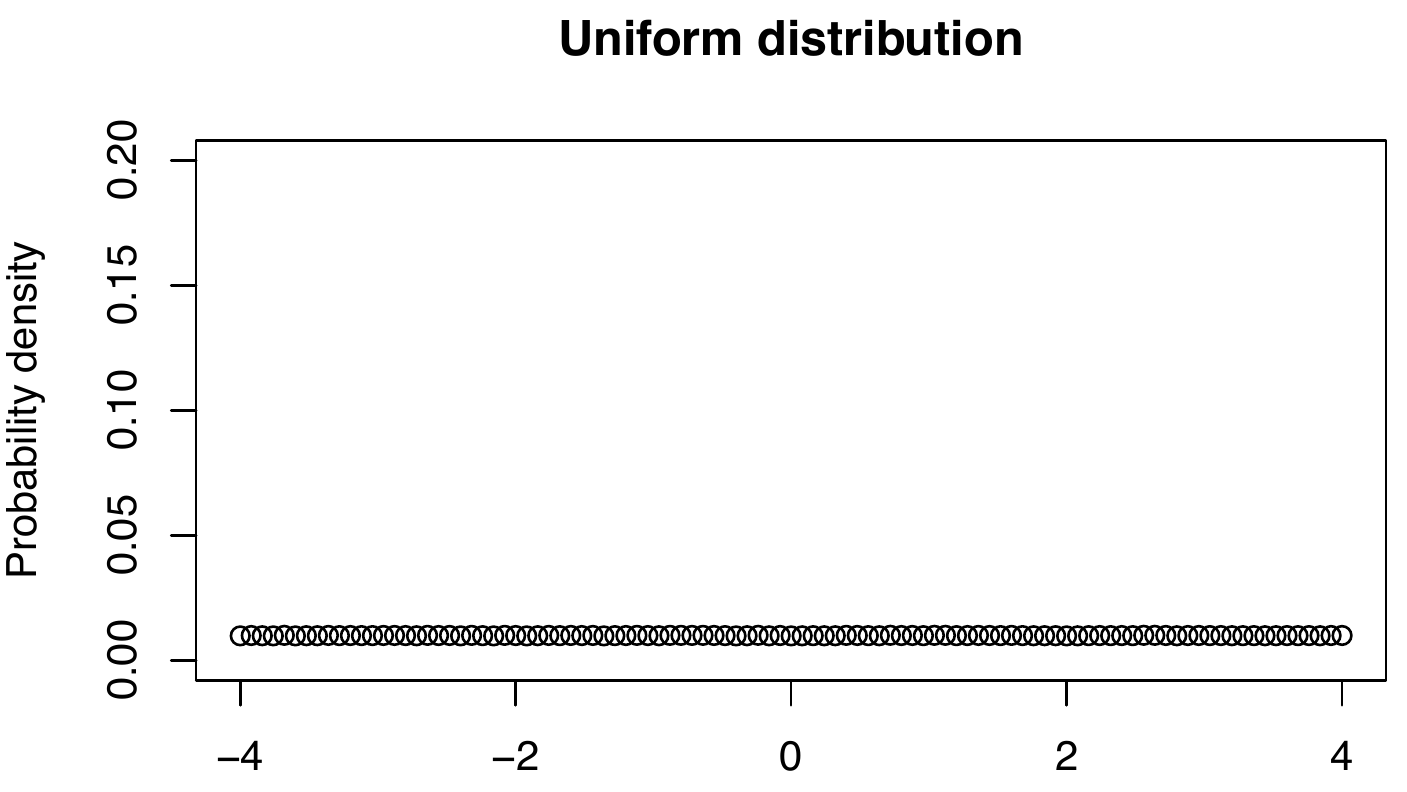}}
\caption{The uniform distribution in the interval $[-4,4]$.\label{fig1}}
\end{figure}

The analytical approach for obtaining this result consists in optimizing the Lagrangian of the continuous entropy of the distribution $f$. The  entropy of $f$ is the negative expected value of the logarithm of $f$:
$$
E= - \int  f(x) {\rm log}(f(x)) dx
$$
Additionally, we have the constraint  $\int f = 1$, so that $f$ represents a probability distribution. The Lagrangian with this constraint is
$$
L(f) = -\int  f(x) ({\rm log}f(x))dx + \lambda \int (f(x)-1)dx
$$
In variational calculus such Lagrangians are optimized solving the Euler-Lagrange equation
$$
\frac{\partial F}{\partial f} - \frac{\partial}{\partial x}\left( \frac{\partial F}{\partial \dot{f}} \right) =0
$$
where $F$ is the function, or sum of functions, inside the integral sign. Since the derivative of $f$ is not present in our Lagrangian, we only have to solve the equation ${\partial F}/{\partial f}=0$. In that case we obtain
$$
-{\rm log}f(x) - 1 + \lambda =0
$$
Since according to the expression above, the logarithm of $f$ is constant, that implies that $f$ itself is constant. It is precisely the uniform distribution. In an interval $[a,b]$, the value of $f$ for any $x$ in $[a,b]$ is $1/(b-a)$. With this value of $f$ the integral of $f$ in the interval $[a,b]$ is 1.

Fig.\ref{fig1} shows the computational result obtained optimizing the Lagrangian numerically, for a discrete distribution with 100 points in the interval [-4,4] following the approach explained in the next section. It is a discrete uniform distribution.

\section{General Optimization Approach}

In general, given the constraints for the optimization problem, a Lagrangian is defined and  we thenfind its extremal values \cite{lisman}. The constraints can be of many types, but usually we have to deal with equality constraints: the mean value has to have a certain value, or the variance, or both. Stating the Lagrangian es straightforward in such cases.

For example, if we require from the distribution $f$ to have the mean value $\mu$ and the variance $\sigma^2$, then the complete Lagrangian, including the constraint $\int f = 1$ is given by:
$$
L(f) = -\int  f(x) {\rm log}(f(x))dx + \lambda_1 \int (f(x)-1)dx + \lambda_2 \int (xf(x)-\mu) dx \\ 
+ \lambda_3 \int ((x-\mu)^2f(x)-\sigma^2) dx
$$ 
Taking the derivative of the functions inside the integrals and setting the result equal to zero we obtain
$$
-{\rm log}(f(x))-1+\lambda_1+ \lambda_2 x + \lambda_3 (x-\mu)^2 =0
$$
This means that the logarithm of $f$ is a quadratic function of $x$, in general, and therefore $f$ is an exponential function with a quadratic function of $x$ in the exponent. Rearranging the Laplace multipliers, we find that the distribution of minimal entropy, given the mean and variance, is a Gaussian distribution.
 
If we want to optimize the above Lagrangian using a numerical approach, there are two alternatives. On the one hand, we can
start with a distribution which fulfills the constraints (in the case above, having a given mean and variance). We then generate a new distribution stochastically, or even several alternative distributions. We pick the one with the highest entropy and continue optimizing. 
When we generate the new distributions, we enforce the constraints by scaling the distribution in an appropriate way. For example, if the variance is too high, we can ``compress'' the distribution  around the mean value in order to reduce the variance.
In this way the optimization procedure never leaves the region of admisible distributions. The worst that can happen is that the evolution of the distributions becomes trapped in a local maximum.

The second alternative is to evolve a given random distribution, generating distorted versions also in a random way. We then pick the best function $f$ in terms of maximizing the Lagrangian and normalizing the result (so that the function is a distribution). Over many iterations, the distribution that numerically maximizes the Lagrangian fulfills all constraints and has maximum entropy.

Of course, in some cases the maximum entropy distribution could not exist for a given set of constraints. In that case the numerical approach will not produce a sensible result, or will not converge.

We have tested both methods numerically and they produce essentially the same results for the examples presented here.

\section{Numerical explorations}

In this section we discuss the discrete distributions obtained by  an evolutionary strategy. We start from a completely random distribution and at each step the probability density is perturbed at a single point.  We experimented with perturbations at several points, without obtaining any significant advantage in terms of convergence speed.

\subsection*{Gaussian distribution}

\begin{figure}[htb]
\centerline{\includegraphics[width=8cm]{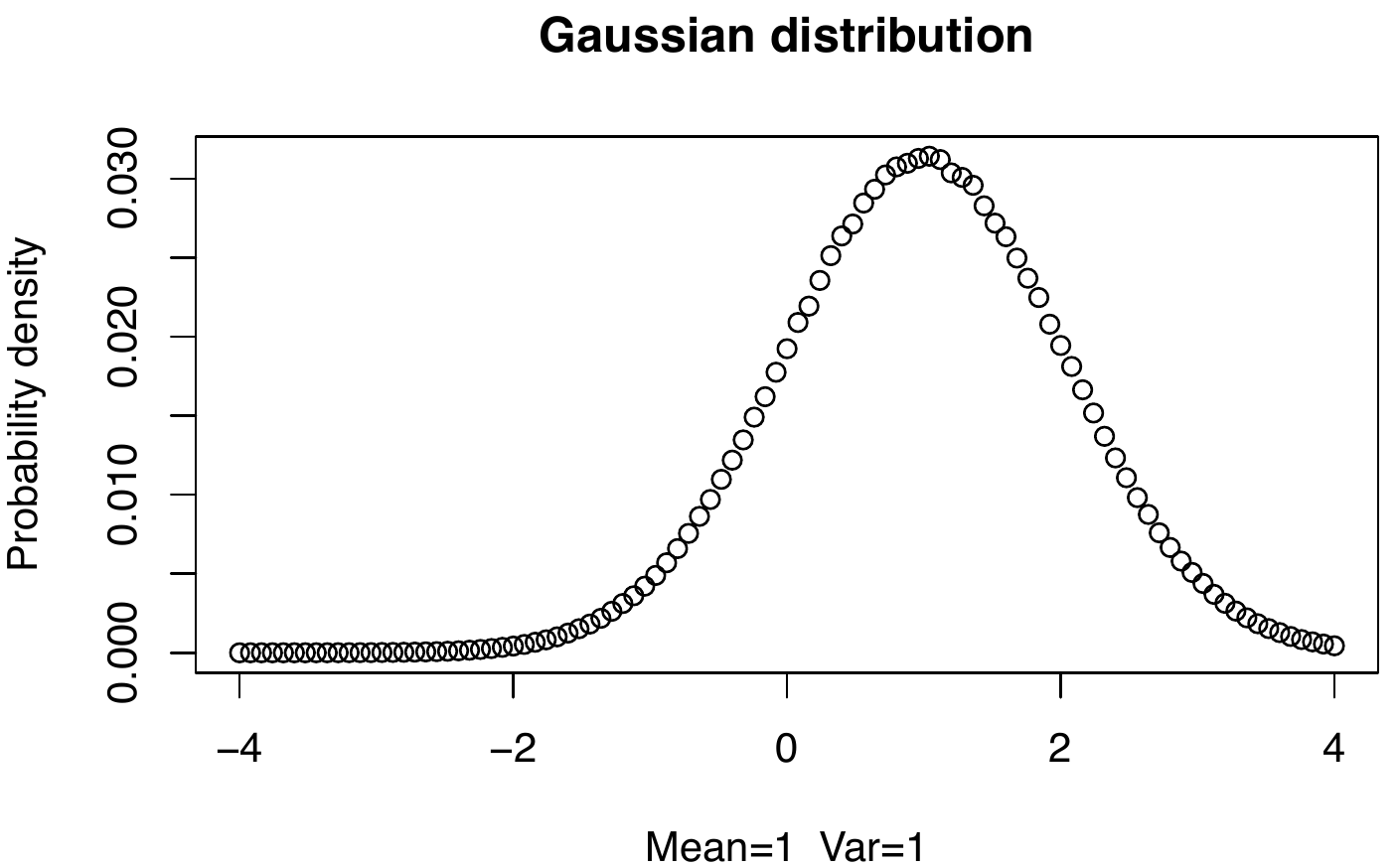}}
\caption{The Gaussian distribution with mean 1 and variance 1.\label{fig2}}
\end{figure}

As we explained above, the Gaussian distribution is obtained when we constraint the value of the mean and of the variance. The constraints are $E[x]=\mu$ and $E[(x-\mu)^2]=\sigma^2$. Including them in the Lagrangian we obtain the distribution shown in Fig.\ref{fig2}.

Notice that the evolutionary procedure is agnostic. The symmetry and continuity of the distribution is not included explicitly in the Lagrangian. We obtain both. Intuitively, this is what we would expect. Given the mean value $\mu$, an asymmetrical distribution would be too special, given that we do not have any information that would give more weight to points to the right or to the left of the mean. The constraint over the variance means that we would expect the distribution to concentrate around the mean value, and that the tails of the distribution should go down asymptotically. Both features of the Gaussian are easy to explain.

A little more difficult to explain is the fact that we obtain a continuous distribution. The reason for this is that in the Langrangian the function $f$ appears multiplying constant terms, or powers of $x$. The derivative ${\partial F}/{\partial f}$ will produce a function involving $\rm{log}f(x)$, powers of $x$ and some constants (some of them the Lagrange multipliers). It is then clear that we can obtain a closed solution, that is, an arithmetical expression for $f(x)$, in terms of powers of $x$.

In the numerical optimization the continuity and symmetry of the function $f$ has not been presupposed. Its appearance is a confirmation that the solution obtained is the distribution of maximum entropy being searched.

\subsection*{Exponential distribution}

\begin{figure}[htb]
\centerline{\includegraphics[width=8cm]{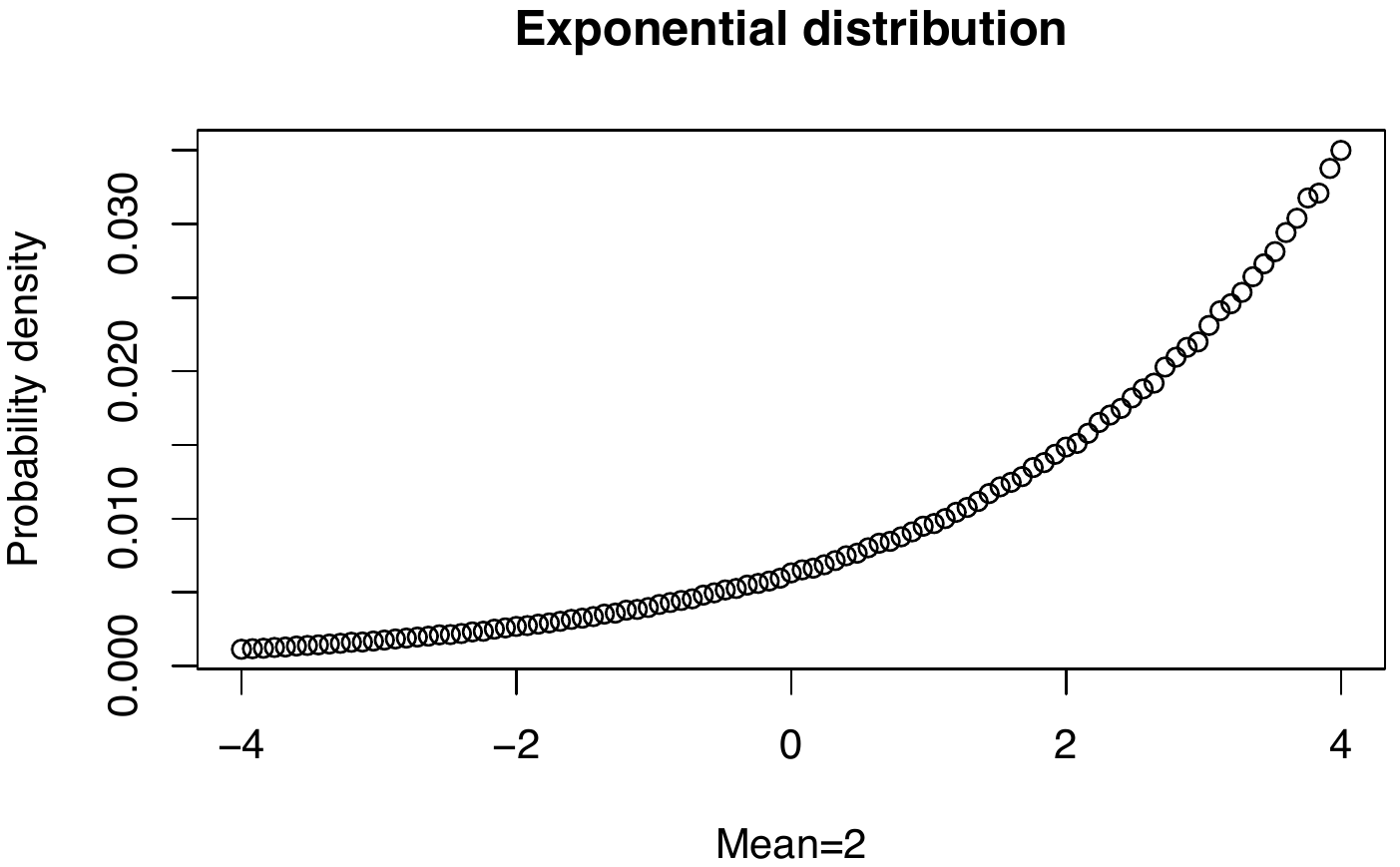}}
\caption{The exponential distribution.\label{fig3}}
\end{figure}

Another interesting result obtained with the numerical approach  described above, is when the only two constraints over $f$ are: a) being a distribution, and b) having a given mean.  Fig.\ref{fig3} shows the  distribution obtained when the mean value has been constrained to be 1. It is interesting to see that we do not obtain a piecewise uniform distribution. We could think that a uniform distribution between -4 and 1, and another, at another level, between 1 and 4 could be optimal. However, the result shows that the distribution tries to cover the two subintervals, but in a continuous manner. Continuity has not been included in the constraints in an explicit way but arises in the way explained in the previous section.
 
We obtain an exponential distribution because the Euler-Lagrange differential equation produces an equation involving  $\rm{log}f(x)$. Solving the equation we obtain and exponential solution for $f(x)$.

If the mean value $\mu$ is located in the middle of the support interval, the exponential distribution becomes ``flat'' and degenerates into a uniform distribution. 

Therefore, if for a given data set we only keep the mean value of the data, and the mean is not in the middle of the support interval, our best guess for the form of the distribution is an exponential function with the shape shown in Fig.\ref{fig3}.

\subsection*{Laplace distribution}

The Laplace distribution is also useful in machine learning. It is obtained when the dispersion of the distribution is measured not by the variance but using the expected value of the absolute deviation from the mean. It is, in some sense, a measure like the variance, but without the square function.

In regression we can measure the deviation of the data points from the regression line using the sum of squared differences. But we could also use the sum of absolute values of the deviations. In that case we obtain a different regression line (the line of least absolute deviation). We use one or the other approach depending on the statistics of the regression error terms.

\begin{figure}[htb]
\centerline{\includegraphics[width=8cm]{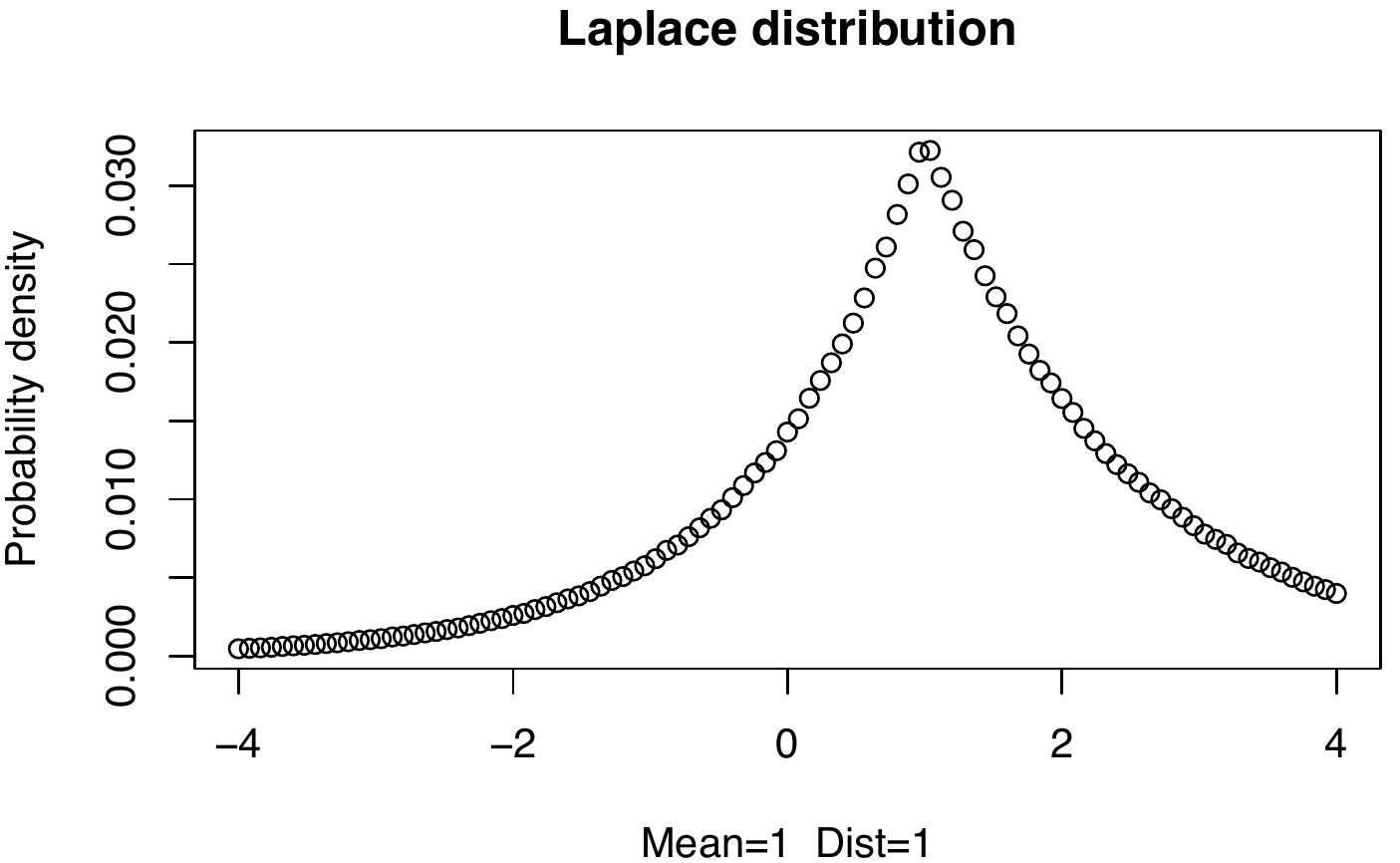}}
\caption{The Laplace distribution.\label{fig4}}
\end{figure}

The Lagrangian for the Laplace distribution is very similar to the Lagrange for the Gaussian distribution. We just have to substitute the square function in the Lagrangian with the absolute value function. We should expect to get a symmetrical and continuous function, as in the case of the Gaussian. The function in Fig.\ref{fig4} shows the numerical result obtained.

The Laplace distribution has been used in machine learning applications. The LASSO regression analysis method can be interpreted as standard regression with a Laplace prior.

\subsection*{Log-normal distribution}

The log-normal distribution is very important in biology, because in may natural phenomena, the effects of successive random effects act multiplicatively instead of additively. A distribution $f$ is log-normal if the logarithm of $f$ has the normal distribution.
Multiplicative effects can be transformed into additive effects by taking the logarithm. 

One example of a process where the log-normal distribution could have an application are growth processes depending on multiple genes. The effect of those genes could be explained assuming that each gene slightly scales up or down an organism. Histograms of the sizes of individuals in a species fit well log-normal distributions.

\begin{figure}[htb]
\centerline{\includegraphics[width=8cm]{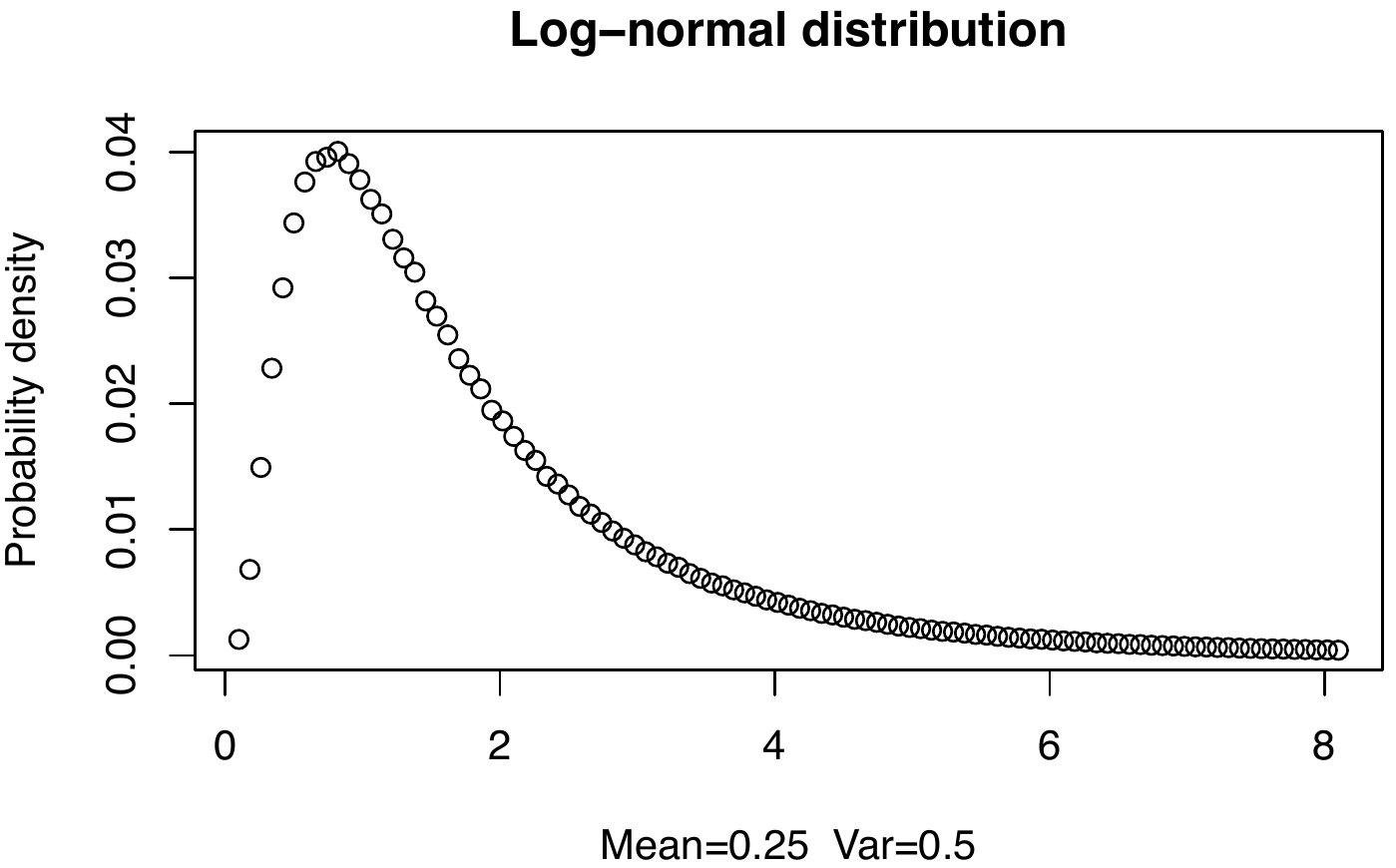}}
\caption{The log-normal distribution\label{fig5}}
\end{figure}

Fig.\ref{fig5} shows the result obtained constraining the mean and variance of $\rm{log} f(x)$.

\subsection*{Median constrained distribution}

The next example is interesting from the point of view that the direct use of the Euler-Lagrange equation is not possible. If we constraint a distribution over a support interval to have a given median, there is no straightforward closed formula that we can insert in the Lagrangian. In this case it is easier to optimize the entropy, selecting only distributions which fulfill the constraint. Given the median, we know that half of the weight of the distribution must be to the left of the median, and the other half to the right. Perturbed distributions can be scaled in order to fulfill the constraint. 

 \begin{figure}[htb]
\centerline{\includegraphics[width=8cm]{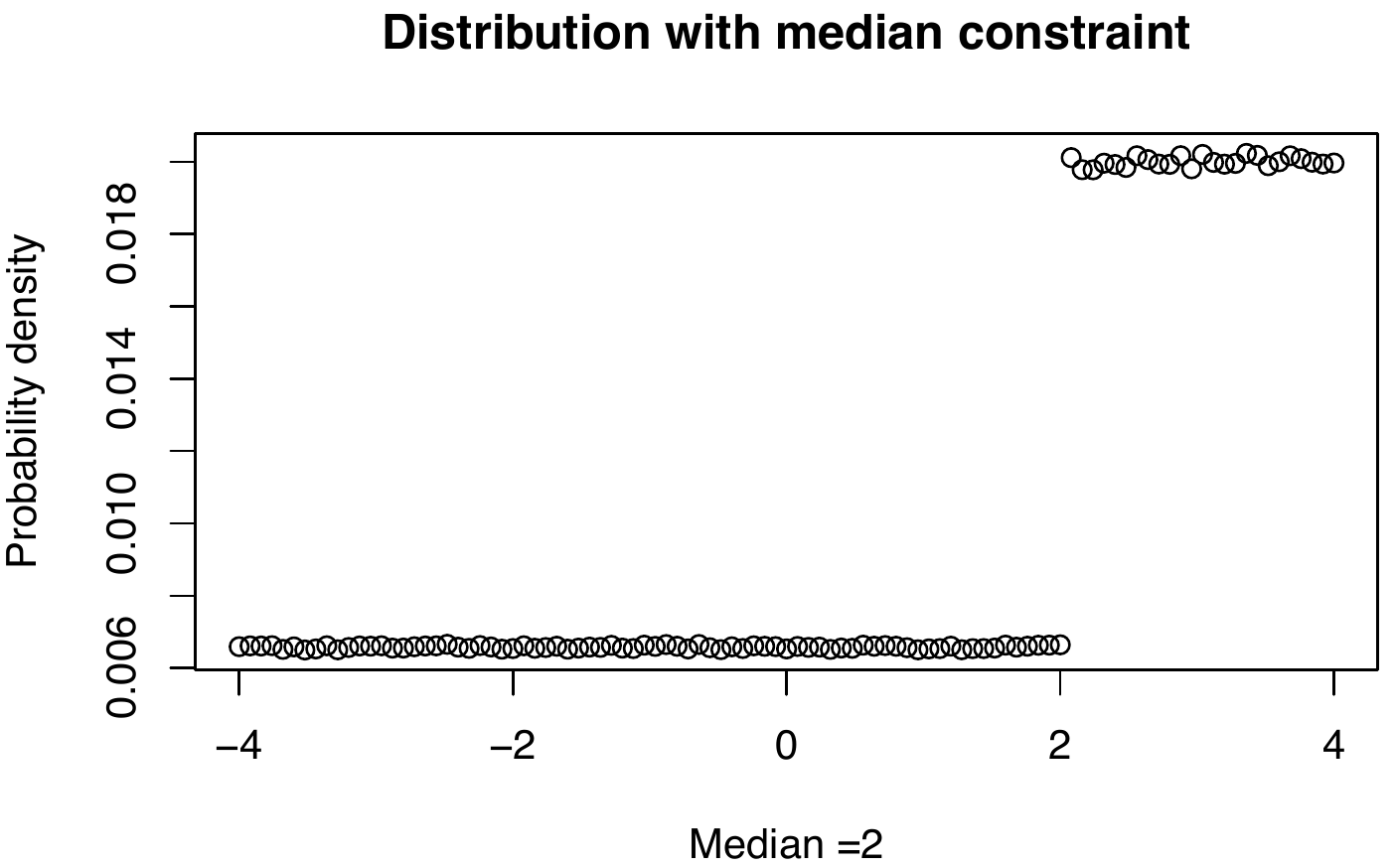}}
\caption{The median constrained distribution\label{fig6}}
\end{figure}

Fig.\ref{fig6} shows the result of the optimization: we obtain two uniform distributions, one to the left and one to the right of the median. It makes sense, since we do not have any other special assumptions about the distribution, other than half of the population being on one side and the other half being on the other side of the median.

\subsection*{Median constrained Laplace and Gaussian distribution}

The next experiment can be easily solved numerically but an analytical solution would be too convoluted. Let us assume that we look for the maximum entropy distributions constrained by a given median, and then either by the expected value of the absolute values of the deviations from the median, or the expected value of the squares of the deviations from the median. I call the first distribution the "median constrained Laplace" distribution, and the second the "median constrained Gaussian".

 \begin{figure}[htb]
\centerline{\includegraphics[width=8cm]{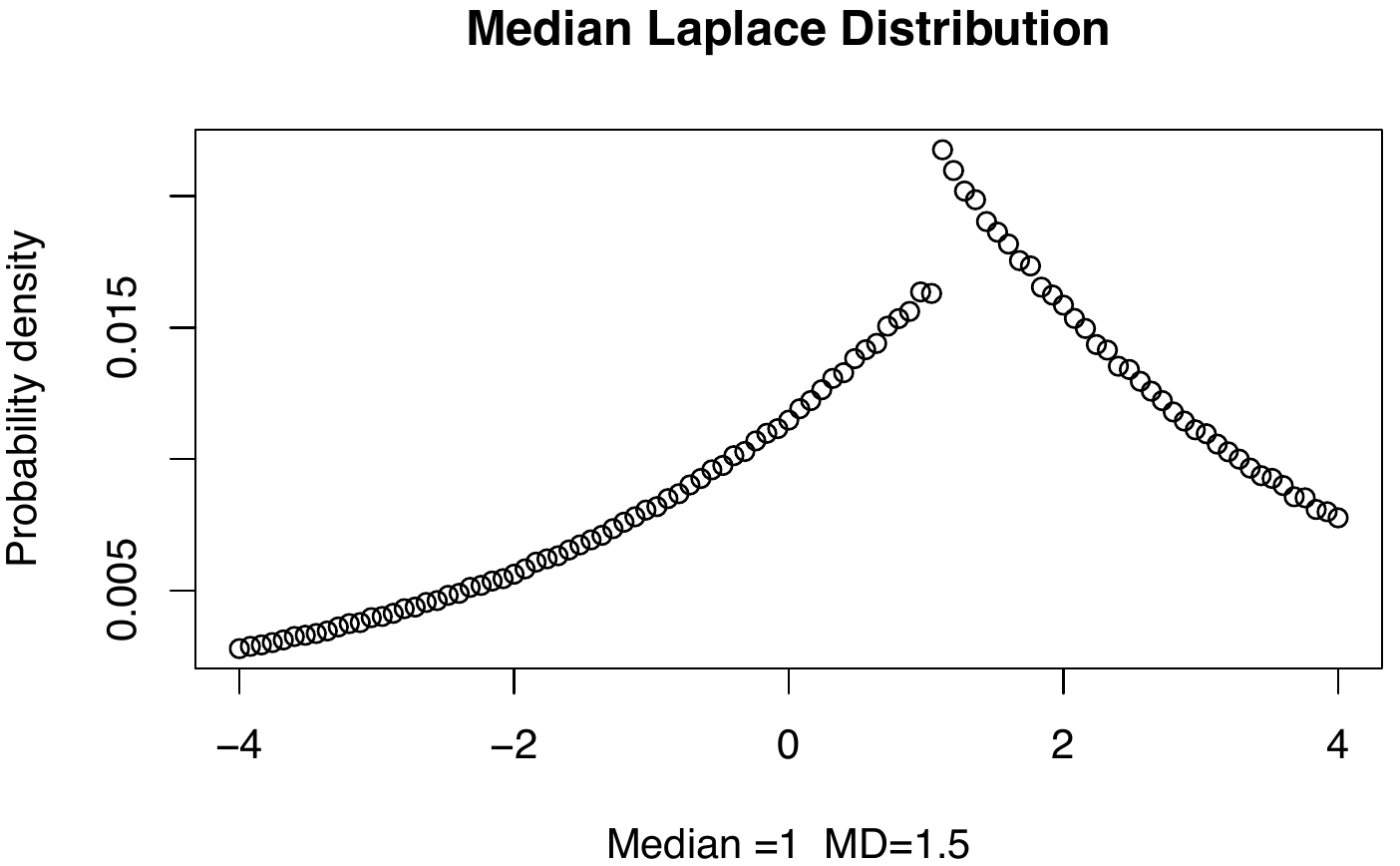}}
\caption{The median constrained Laplace distribution\label{fig7}}
\end{figure}

Fig.\ref{fig7} shows the shape of the distribution with a given median and an expected value of the distance to the median with value 1. Now we lose the continuity of the complete distribution, because the median constrain has the effect of dividing the support  interval into two disconnected compartments. Half of the total probability is on one side of the median, half of the probability on the other side. The constrain over the expected absolute deviation can be fulfilled with maximum entropy solving two actually disjoint problems. The distribution curves do not touch at the median.

 \begin{figure}[htb]
\centerline{\includegraphics[width=8cm]{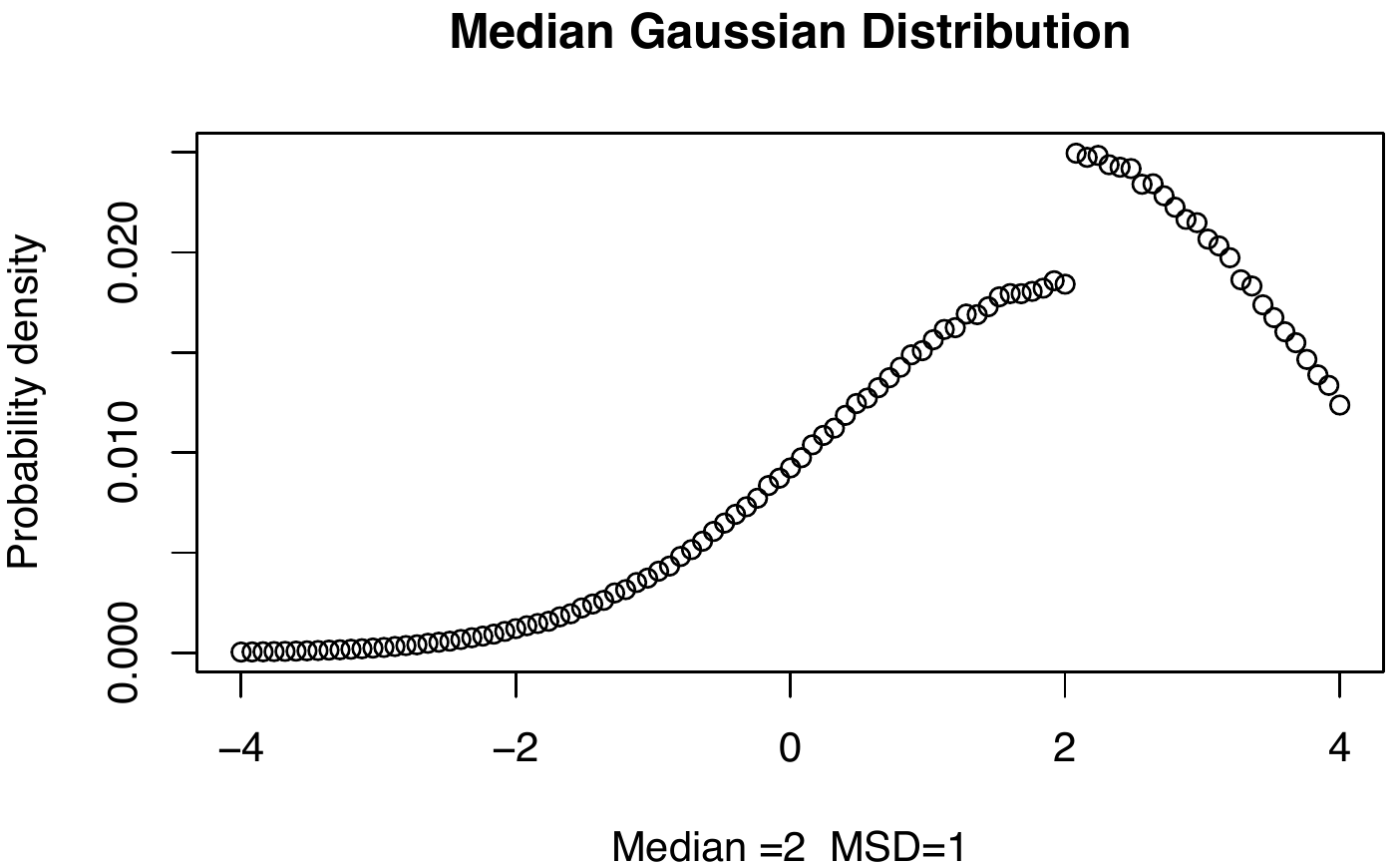}}
\caption{The median constrained Gaussian \label{fig8}}
\end{figure}

The same happens in the case of Fig.\ref{fig8}, where we see the "median Gaussian". The result is equivalent to looking for maximum entropy distributions with a given variance to the left and right of the median.

It would be interesting to think about applications where such median constrained distributions could make sense. Here, they are mentioned as  interesting examples of cases in which direct optimization of the Lagrangian for a subset of distributions that fulfill all or some of the contraints leads directly  to the solution. In the two cases presented in this section, the median constrained was enforced directly on the generated distributions, while the maximum entropy and expected value of the deviations was left in the Lagrangian.

\subsection*{Chimera distributions}

\begin{figure}[htb]
\centerline{\includegraphics[width=7cm]{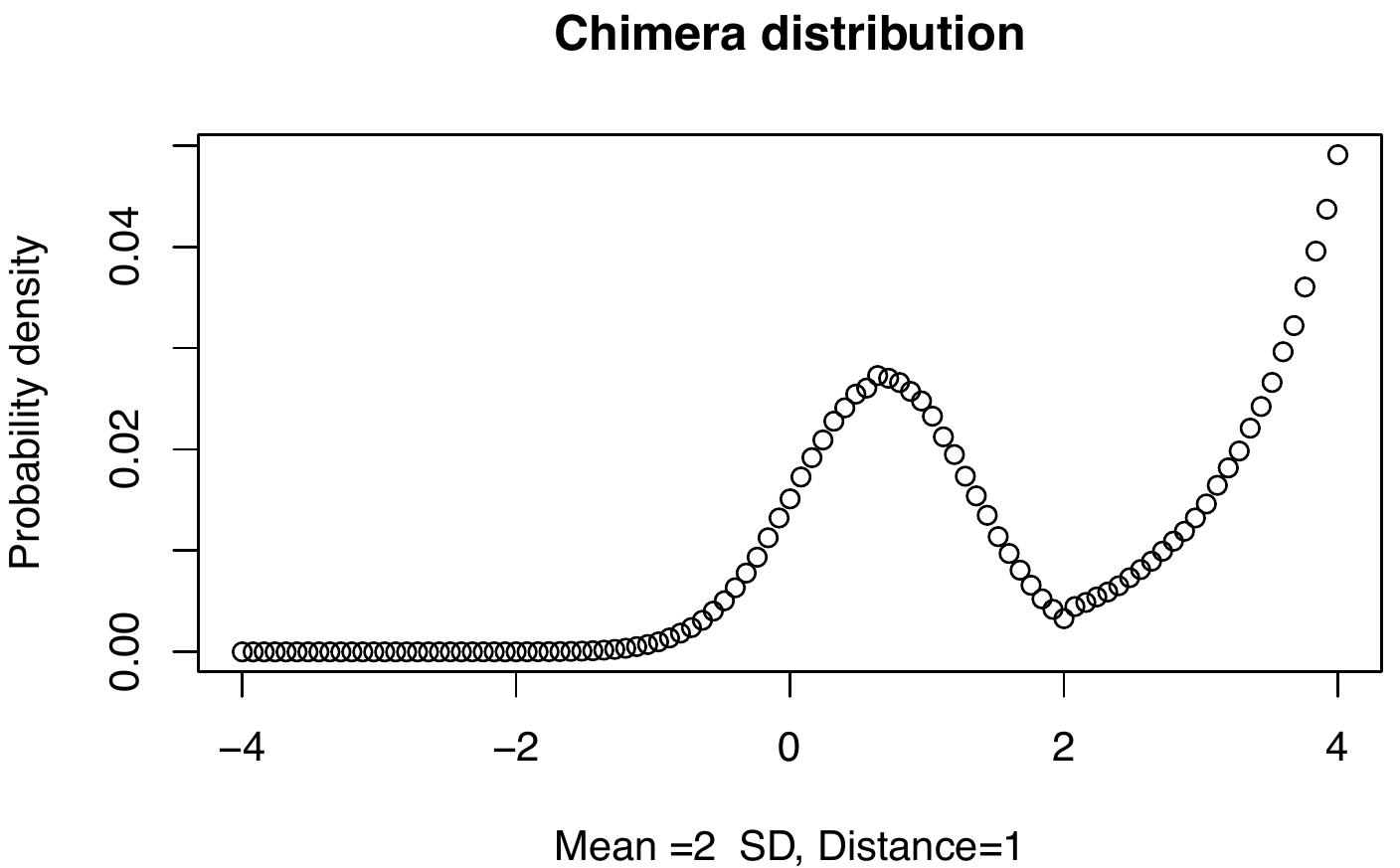}}
\centerline{\includegraphics[width=7cm]{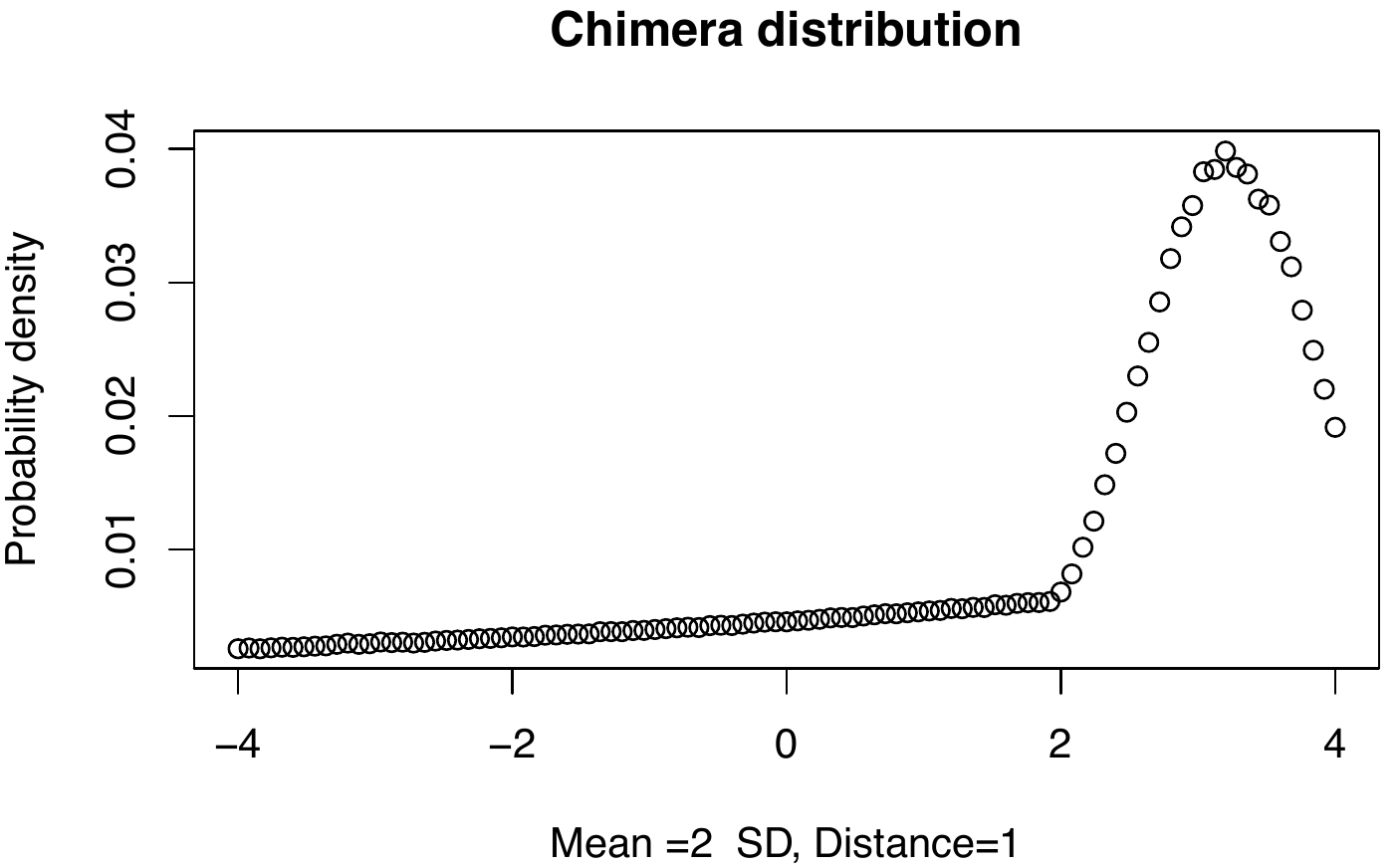}}
\caption{A half Laplace, half normal distribution.\label{fig9}}
\end{figure}

We can become bolder now and investigate "chimera distributions", that is, combinations of two different distributions.
An example could be a generalization of skewed distributions with tails around the mean with different shapes.
 We can, for example, look for the maximum entropy distribution with a given mean, but where on the left of the distribution we constrain the expected absolute deviation from the mean, while on the right we constrain the expected squared deviation from the mean. The results are shown in Fig.\ref{fig9}. In the first case, on the left side we have a Gaussian while we have a Laplace distribution on the right. In the second case, the
 Laplacian and Gaussian sides have been transposed.

As can be seen in the figures, the chimera distributions show some kind of continuity at the interface with the mean.
This is not necessarily so. Additional experiments with different numerical constraints show that the two pieces of the distribution can disconnect at the mean. It would be interesting to investigate under which circumstances the resulting chimera distribution can be continuous.

\subsection*{Cauchy distribution}

\begin{figure}[htb]
\centerline{\includegraphics[width=8cm]{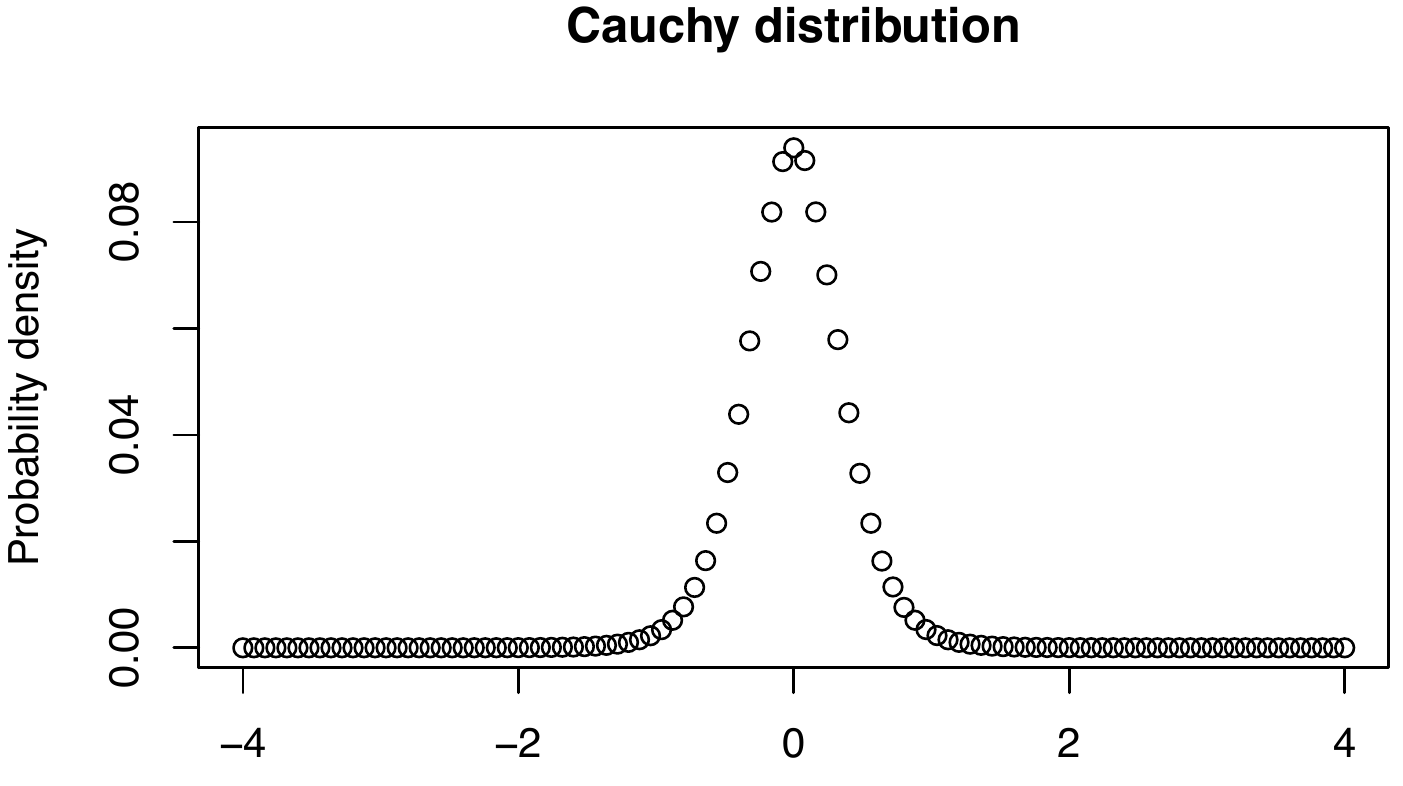}}
\caption{The Cauchy distribution.\label{fig11}}
\end{figure}

Another example is the Cauchy distribution which does not have finite moments of order greater than one, but which nevertheless is the distribution of maximum entropy
when the expected value of $\rm{log}(1+x^2)$ is constrained to be a certain constant. The Cauchy distribution describes the distribution of the ratio of two normally distributed random variables and has applications when describing spinning objects.
Fig.\ref{fig11} shows the shape of the Cauchy distribution obtained numerically.

\subsection*{Chi-squared distribution}

\begin{figure}[htb]
\centerline{\includegraphics[width=8cm]{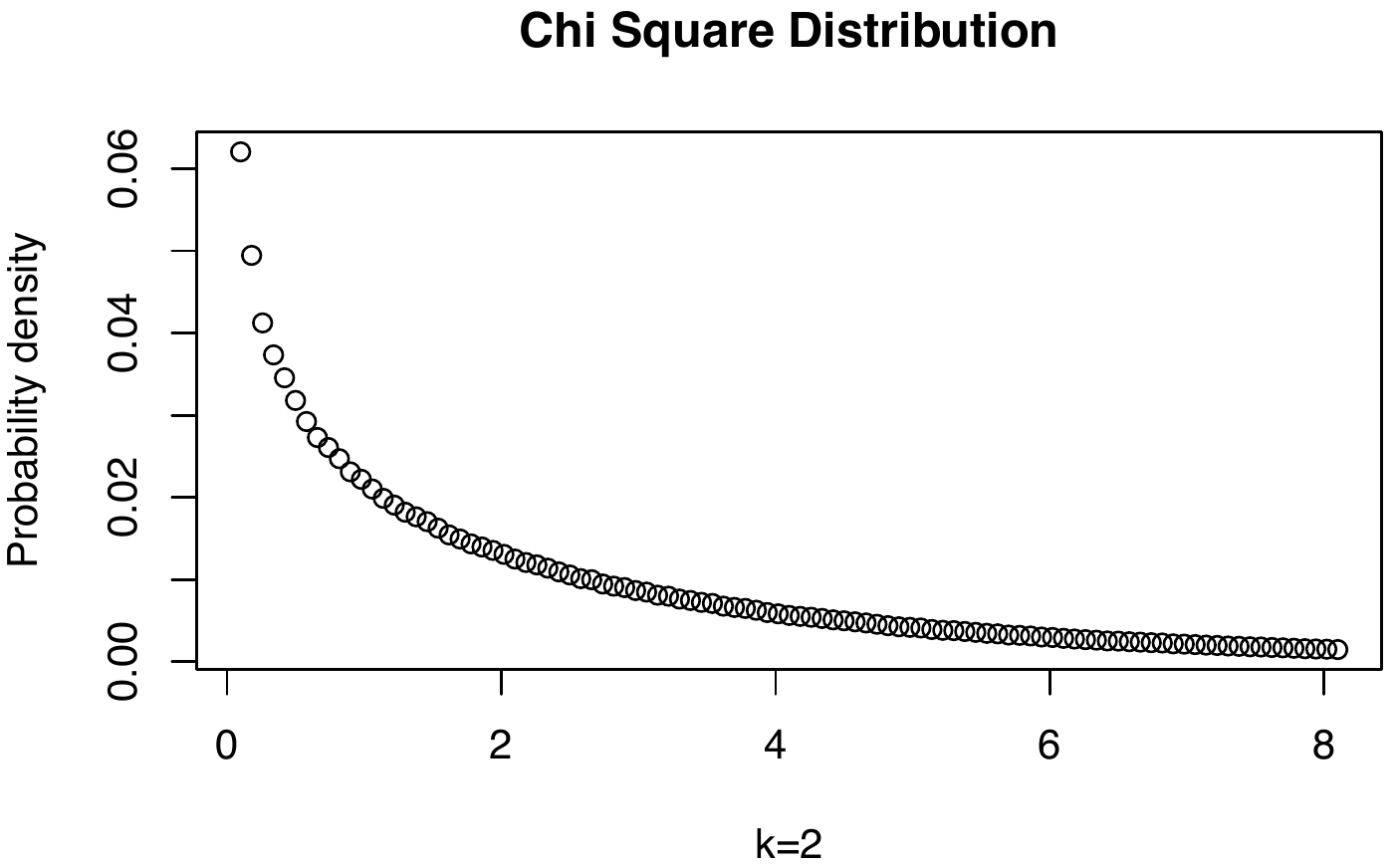}}
\caption{The Chi-squared distribution.\label{fig12}}
\end{figure}

The Chi-squared distribution is the distribution of the sum of the squares of $k$ independent standard normal variables.
In the example handled here, the support has been constrained to the interval $[0,8]$, while usually it is unbounded to the right.
Fig.\ref{fig12} shows the shape of the Chi-squared distribution obtained numerically. It differs from the Chi-squared distribution over
the unbounded interval $[0,\infty]$ because of the compact support.

\section{Conclusions}

This paper has shown that maximum entropy distributions can be readily found once the constraints over the distribution are
inserted into a Lagrangian to be optimized. For the optimization an approach that evolves distributions numerically from an initial, randomly chosen distribution, can be used. We showed that some classical analytical results can be reproduced for discrete distributions, but also that in the case that an analytical approach is not possible or just too difficult, the numerical approach can be
a good exploratory tool. It can be used directly in some applications \cite{buck}.

In the numerical examples illustrated here, we can use a Lagrangian which only includes the entropy function, enforcing the constraints during the evolutionary process, or we can have some constraints in the Lagrangian while others are enforced in the evolutionary process. In the case of distributions with a constraint over the median, it is easier to enforce the median constraint in the evolutionary process and leave the other constraints in the Lagrangian.

We have also defined "chimera" distributions in this paper as those subject to different constraints over the support interval. In particular we have presented the Laplace-Gaussian skewed  distribution. It would be interesting to investigate in which applications such chimera distributions could be useful.

For educational purposes, it is also interesting that the evolutionary process can be visualized while the constraints are introduced step by step. We can start from distribution with no constraints and then displace the mean to one side of the support interval. The distribution gradually transforms into an exponential. We can then add constraints over the variance and the distribution has to bend down in order to keep the variance in check. Movies of this gradual procedure can give students a good feeling for the way in which the maximum entropy principle constrains the shape of the distribution.


\begin{thebibliography}{999999}

\bibitem[Cover 06]{cover} Th. Cover, {\it Elements of Information Theory}, Wiley, 2006.
\bibitem[Rojas 96]{rojas} R. Rojas, {\it Neural Networks}, Springer-Verlag, 1996.
\bibitem[Lisman 72]{lisman} Lisman, J. H. C., van Zuylen, M. C. A., {"Note on the generation of most probable frequency distributions"}, {\it Statistica Neerlandica}, 26 (1): 19–23, 1972.
\bibitem[Buck 91]{buck}Brian Buck, Vincent A. Macaulay (eds), {\it Maximum Entropy in Action: A Collection of Expository Essays}, Oxford University Press, Oxford, 1991. 

\end{thebibliography}
\end{document}